\documentclass[aps,prl,twocolumn,superscriptaddress,preprintnumbers]{revtex4-2}
\pdfoutput=1
\usepackage{amsmath}
\usepackage{graphicx}
\usepackage{bbm}
\usepackage{amssymb}
\usepackage{enumitem}
\usepackage{booktabs}
\usepackage{verbatim} 
\usepackage{soul}
\usepackage{mathrsfs}
\usepackage[linktocpage=true,
colorlinks=true,
urlcolor=blue,
anchorcolor=blue,
citecolor=blue,
filecolor=blue,
linkcolor=blue,
menucolor=blue,
]{hyperref}

\newcommand{\diff}{\mathrm{d}}
\newcommand{\dd}{\mathrm{d}}
\newcommand{\R}{\mathbb{R}}
\newcommand{\vol}{\mathrm{vol}}

\newcommand{\hook}{\mathbin{\rule[.2ex]{.4em}{.03em}\rule[.2ex]{.03em}{.9ex}}}

\usepackage{color}

\def\nn{\nonumber}
\newcommand{\ii}{\mathrm{i}}

\newcommand{\Z}{\mathbb{Z}}

\newcommand{\Deltathere}{\Delta_{\mathrm{there}}}

\newcommand{\newOmega}{\Upsilon}

\newcommand{\wwlog}{w}

\newcommand{\Phiprime}{\Phi'}
\newcommand{\wprime}{w'}

\begin{document}

\title{
Equivariant localization for higher derivative supergravity}

\author{Pietro Benetti Genolini}
\affiliation{D\'epartement de Physique Th\'eorique, Universit\'e de Gen\`eve, 24 quai Ernest-Ansermet, 1211 Gen\`eve, Suisse}

\author{Florian Gaar}
\affiliation{Mathematical Institute, University of Oxford, Woodstock Road, Oxford, OX2 6GG, U.K.}

\author{Jerome P. Gauntlett}
\affiliation{Blackett Laboratory, Imperial College, Prince Consort Road, London, SW7 2AZ, U.K.}

\author{James Sparks}
\affiliation{Mathematical Institute, University of Oxford, Woodstock Road, Oxford, OX2 6GG, U.K.}

\begin{abstract}
\noindent 
Conformal supergravity provides 
an effective
off-shell formalism to study  higher derivative actions. We show that the $D=4$, $\mathcal{N}=2$ 
theory admits equivariantly closed forms. These may be used to compute 
closed-form expressions for supersymmetric observables 
in a general class of supergravity theories with higher derivative couplings, without any need to  solve equations of motion. We discuss applications to holography, presenting results for on-shell actions that are conjecturally valid to all orders in the
perturbative $1/N$ expansion.

\end{abstract}

\maketitle

\enlargethispage{0.25\baselineskip}

\section{Introduction}\label{sec:intro}
Equivariant localization is a powerful new technique in supergravity \cite{BenettiGenolini:2023kxp}. 
It allows one to compute various supersymmetric observables, 
including on-shell actions, black hole entropies and central charges, 
without solving equations of motion. 
The structure seems to be universal, applying to supergravity theories coupled to arbitrary matter
in general dimensions. However, so far the theory has only been developed 
for two-derivative theories.

On the other hand, higher derivative corrections to supergravity 
play an important role in our understanding of quantum gravity. 
This includes the computation of all-order corrections to the 
Bekenstein--Hawking--Wald 
entropy of supersymmetric asymptotically flat black holes, matching to a microstate counting, as reviewed in \cite{Mohaupt:2000mj, Mandal:2010cj}. More recently,
higher derivative corrections to gauged supergravity have led to new 
tests of holography \cite{Bobev:2020egg, Bobev:2021oku,Genolini:2021urf}, with 
various conjectures for generalizations of these results 
 to all derivatives and coupling to matter 
 appearing \cite{Bobev:2021oku,Hristov:2021qsw, Hristov:2022lcw, Bobev:2022eus, Hristov:2024cgj}.

An obstacle in extending any higher derivative program is that, in general, constructing explicit supersymmetric solutions is intractable. 
The Lagrangians are complicated, and solving the resulting 
equations of motion is a formidable task even for solutions with a high degree of supersymmetry/symmetry. 
 Here we show that this obstacle can be bypassed by extending the localization techniques of \cite{BenettiGenolini:2023kxp} to conformal supergravity. This allows us to analyse
a broad class of all-order higher derivative couplings with vector multiplet matter in $D=4$, $\smash{\mathcal{N}=2}$ off-shell supergravity. 

The approach contains three main ingredients. 
First, a 
conformal Killing vector field $\xi$ that can be constructed as a bilinear in the Killing spinor parametrizing the preserved supersymmetry. Second, a set of polyforms 
built from Killing spinor bilinears and supergravity fields that are annihilated by the differential operator $\smash{\dd_\xi \equiv \dd - \xi \hook\mskip2mu}$. 
Third, localization theorems \cite{BV:1982,Atiyah:1984px} 
and supersymmetry allow one to evaluate integrals of these differential forms purely in terms of the fixed point set where $\xi=0$.
These integrals may then be used to compute physical observables of the supersymmetric configuration, including the action, in terms of global topological data.

Our results allow us to prove various aspects of the holographic conjectures \cite{Bobev:2021oku,Hristov:2021qsw, Hristov:2022lcw, Bobev:2022eus, Hristov:2024cgj} and also extend them, thereby providing new tools 
in supersymmetric holography. 
They also apply to ungauged supergravity, with applications to asymptotically flat solutions.
Analogous equivariant structures in conformal supergravity in other 
dimensions should also exist, with a variety of applications. Additional results for the $D=4$ case appear
in \cite{toappone} and a localization result for $D=5$ $AdS$ black holes appears in \cite{toapptwo}.

\section{Conformal supergravity}
Higher derivative terms in 
supergravity are most easily constructed by starting with conformal supergravity, which realises supersymmetry off-shell. 
We consider $D=4$, $\mathcal{N}=2$ conformal supergravity in Euclidean signature \cite{deWit:2017cle}. The theory is
obtained by gauging the superconformal group
whose bosonic group consists of the product of the Euclidean conformal group, $SO(5,1)$, and the R-symmetry group $SU(2)\times SO(1,1)$. 
Various multiplets may be constructed, with a proliferation 
of auxiliary fields required to close the superconformal algebra.
For our purposes only a few key combinations 
of fields are important.

The  basic multiplet of the theory is the Weyl multiplet, which is associated with gauging the superconformal algebra.  
 The bosonic fields are
 \begin{align}
\text{Weyl:}\quad   \{e_\mu^a, b_\mu, \mathcal{V}_{\mu j}^i, A_\mu,\omega_\mu^{ab}, f_\mu^a, T_{ab}, D\}\, .
\end{align}
There are gauge fields associated with
general coordinate transformations, namely the orthonormal frame $e^a_\mu$, 
  dilatations, $b_\mu$, and the $SU(2)\times SO(1, 1)$ R-symmetry,
$\mathcal{V}_{\mu j}^i, A_\mu$. 
Here $\mu,\nu =1,\dots,4$ and $a,b=1,\dots 4$ are spacetime and frame indices, while $i,j=1,2$ are $SU(2)$ R-symmetry indices. 
The multiplet includes composite gauge fields $\omega_\mu^{ab}, f_\mu^a$ and 
 auxiliary fields $T_{ab},D$, as well as fermions, which we suppress. 
 The anti-symmetric tensor $T_{ab}$ will feature prominently.  
 The bosons are all real, except for $\mathcal{V}_{\mu j}^i$ which is anti-Hermitian.
Since we are ultimately interested in Poincar\'e supergravity,  we partially gauge fix 
by setting $b_\mu=0$, as is common.

Next, we consider
chiral and anti-chiral multiplets, $\Phi_\pm$, with the $\pm$ sign correlated with chirality.
For brevity, we refer to
both of these as ``chiral multiplets.'' The bosonic 
fields are
\begin{align}\label{chiralcomponents}
\text{Chiral $\Phi_\pm$:}\quad   \{A_\pm,B^{ij}_\pm,F^{\mp}_{ab},C_\pm  \}\, .
\end{align}
Here $A_\pm$, $C_\pm$ are real scalar fields,
$F^\mp_{ab}$ are real anti-self-dual/self-dual two-forms,
$F^\mp_{ab}=\mp\frac{1}{2!}\varepsilon_{ab}{}^{cd}F^\mp_{cd}$, and
$B^{ij}_\pm=B^{(ij)}_\pm$ and satisfies the reality condition $(B^{ij}_\pm)^*=
 \varepsilon_{ik}\varepsilon_{jl} B^{kl}_\pm$.
All fields in the multiplet transform under Weyl 
transformations 
with a particular weight. For example, if $A_\pm$ has weight $w$ then $C_\pm$ has weight $w+2$, and 
the multiplet is then said to have weight $w$. The product of two chiral (or two anti-chiral) multiplets of weights $w_1$ and $w_2$ yields a chiral (or anti-chiral) multiplet of weight $w_1+w_2$, with components given in \cite{deWit:2017cle}. 
For example, the Weyl multiplet is itself the constrained sum of chiral and anti-chiral tensor multiplets;
using them we can construct scalar chiral and anti-chiral Weyl-squared multiplets, $\mathbb{W}^2_\pm$,
of weight $w=2$,
whose components 
are given in \cite{deWit:2017cle}.

The bosonic fields of a single vector multiplet $\mathbb{V}$ are  
\begin{align}
\text{Vector $\mathbb{V}$:}\quad \{X_\pm, W_\mu, Y^{ij}   \}\,.
\end{align}
Here $X_\pm$ are two real scalar fields with weight $w=1$,
$W_\mu$ is a real gauge field and 
$Y^{ij}=Y^{(ij)}$ satisfies 
$(Y^{ij})^*
= \varepsilon_{ik}\varepsilon_{jl} Y^{kl}$.
For multiple vector multiplets we 
append an index $I$.
We focus on the case of Abelian multiplets for simplicity and write $F(W)=\dd W$ for the curvature. 
A vector multiplet is a constrained combination of a chiral and an anti-chiral multiplet each with weight $w=1$;
expressions for the components are given in \cite{deWit:2017cle}. 

Finally, recall the T-log multiplet \cite{Butter:2013lta}. The construction starts with an anti-chiral/chiral multiplet $\Phiprime_\mp$ of weight $\wprime\ne 0$ in superspace. 
One then constructs the logarithm and applies 
four superspace derivatives to obtain a chiral/anti-chiral T-log multiplet of
weight $w=2$, denoted ${\mathbb T}_\pm$.

Chiral and anti-chiral multiplets of weight $w=2$ may be used to construct 
supersymmetric 
 invariant actions, with Lagrangian four-forms $\mathcal{L}_\pm$ of weight $w=4$. 
The bosonic part is given by 
\begin{align}\label{genactform}
I_\pm = \int \mathcal{L}_\pm = \int \frac{1}{2}\Big[C_\pm +\frac{1}{16}A_\pm (T^\pm)^2\Big]\vol_4\,,
\end{align}
where $(T^\pm)^2\equiv T^\pm_{ab}T^{\pm ab}$,
and $\vol_4$ denotes the usual
volume form. Each multiplet above forms 
an off-shell
representation of the superconformal algebra, i.e. without needing 
to impose equations of motion from an action. 

\section{Equivariantly closed forms}\label{sec:eqclosed}
We are interested in \emph{supersymmetric configurations} of off-shell conformal supergravity, which are defined to be bosonic
configurations such that the corresponding supersymmetry transformations vanish. These transformations 
are parametrized by symplectic Majorana spinors 
$\epsilon^i$ and $\eta^i$, corresponding to the $\mathcal{N}=2$ $Q$- and $S$-supersymmetries, respectively \footnote{The symplectic Majorana condition on a spinor $\epsilon^i$ is $\epsilon_i^* = C\varepsilon_{ij}\epsilon^j$, where
 $C$ is the charge conjugation matrix.}. 
Only the $Q$-supersymmetry parameters will appear in what follows, 
with chiral components $\epsilon^i_\pm\equiv 
\tfrac{1}{2}(1\pm\gamma_5)\epsilon^i$, where 
$\gamma_5$ is the chirality operator, while the $\eta^i$
have been eliminated.

Given a supersymmetric configuration with spinor $\epsilon^i$, 
we introduce the following 
real scalars, real one-form and complex two-form bilinears:
\begin{align}\label{spmdef}
S_\pm&\equiv \bar\epsilon_{i\pm}\epsilon^i_\pm \, ,\qquad 
\xi^\flat\equiv 2\ii \bar\epsilon_{i+}\gamma_{(1)}\epsilon^i_-  \,, 
\nonumber\\  
\newOmega_{\pm i}{}^{j}& \equiv \bar\epsilon_{i\pm}\gamma_{(2)}\epsilon^j_\pm\, .
\end{align}
Here $\gamma_\mu=\gamma_\mu^\dagger$, $\gamma_{(r)}\equiv\tfrac{1}{r!}\gamma_{\mu_1\cdots\mu_r}\diff x^{\mu_1}\wedge\cdots\wedge \diff x^{\mu_r}$
and $\bar\epsilon\equiv \epsilon^\dagger$. 
The supersymmetry transformations imply that the vector $\xi^\mu$, dual to the one-form $\xi^\flat_\mu$, is a conformal Killing vector 
\cite{Klare:2013dka}. Furthermore,
since $\|\xi^\flat\|^2=4S_+S_-$ this vector has fixed points when either $S_+ =0$ or $S_-=0$, 
and at such a fixed point, the spinor $\epsilon^i$ will be of definite chirality $\epsilon^i_-$ or $\epsilon^i_+$,
respectively.

Following \cite{BenettiGenolini:2023kxp}, we next construct polyforms $\Psi$ 
out of the bilinears \eqref{spmdef} and supergravity fields, 
which are closed under the equivariant exterior derivative $\dd_\xi \equiv \dd-\xi\hook\mskip2mu$, so $\dd_\xi \Psi=0$. 
Starting from the Weyl multiplet and a \emph{general} chiral multiplet 
$\Phi_\pm$  of weight $w=2$, with components given in
\eqref{chiralcomponents}, we find that $\Psi^\pm=\Psi^\pm_4+\Psi^\pm_2+\Psi^\pm_0$ 
are equivariantly closed, where
\begin{align}\label{eqclosedactwo}
\Psi^\pm_4&\equiv  \frac{1}{2}\Big[C_\pm +\frac{1}{16}A_\pm (T^\pm)^2\Big]\vol_4
\, ,\nn\\
\Psi^\pm_2  &\equiv
2S_\mp\big(F^\mp + \frac{1}{2} A_\pm T^\pm\big)  -
  B^{ij}_\pm\varepsilon_{jk}\newOmega_{\mp i}^{\ \ \ k}\, ,\nn\\
\Psi^\pm_0  &\equiv\pm 4A_\pm S_\mp^2\,.
\end{align}
Note that $\Psi_4^\pm$ is the Lagrangian four-form in  \eqref{genactform}; 
we emphasize that 
$\dd_\xi \Psi^\pm =0$ follows from
off-shell supersymmetry, without imposing equations of motion that follow from a Lagrangian.

Next, for each vector multiplet 
we can construct an additional equivariantly closed form
\begin{align}\label{fieldstrengthec}
	\Psi_{(F)}&\equiv F(W) +2(X_+S_--X_-S_+)\,,
	\end{align}
	where $F(W)$ is the field strength.
This plays an important role in deriving the key ``gluing rules'', below.

To evaluate $\Psi_0^\pm$ in \eqref{eqclosedactwo} at fixed points 
of $\xi$, below, we need further input from supersymmetry. 
Starting from a general chiral 
multiplet $\Phiprime_\pm$ of arbitrary weight $\wprime$, we 
obtain the expression \eqref{dxieqappendix} for the two-form $\dd\xi^\flat$. 
At a fixed point where $\xi=0$,  the skew eigenvalues of $\tfrac{1}{2}\dd\xi^\flat$ define
the weights $b_i$, $i=1,2$, of the action of the vector field $\xi$. The connected components of the fixed point set 
are either isolated points (``nuts") or surfaces (``bolts") \cite{Gibbons:1979xm}.
At a nut, $b_1, b_2\ne 0$ and we can locally write $\xi=\sum_{i=1}^2b_i\partial_{\varphi_i}$ where $\partial_{\varphi_i}$ rotate each of the 
$\mathbb{R}^2_i$ in the tangent space $\mathbb{R}^4=\mathbb{R}^2_1\oplus \mathbb{R}^2_2$ at the nut. 
At a bolt, which will be some fixed Riemann surface, 
 either $b_1=0$ or $b_2=0$, with $b\mskip1mu \partial_{\varphi}$ rotating the 
normal $\R^2$. 
A computation using \eqref{dxieqappendix} 
reveals that at a $\mp$ chirality fixed point 
we have 
\begin{align}\label{b1pmb2}
	(b_1\mp b_2)^2 &
	=\frac{1}{16}S_\mp^2 A_{\pm}|_{\mathbb{W}^2_\pm}\,,\,
	\nn\\
	(b_1\pm b_2)^2&=	\frac{2}{\wprime}S_\mp^2 A_\pm|_{\mathbb{T}_\pm}\,,
\end{align}
where $A_{\pm}|_{\mathbb{W}^2_\pm}=(T^\mp)^2$ and $A_\pm|_{\mathbb{T}_\pm}$ given by \eqref{tlogaexpressions} are the lowest weight components of the
Weyl-squared multiplet and the T-log multiplet, respectively. 
Equations \eqref{b1pmb2} play a crucial role in localizing higher derivatives.

\section{Localization}
We can now evaluate the action \eqref{genactform} for a general off-shell 
supersymmetric configuration, for an arbitrary chiral or anti-chiral multiplet of weight $w=2$. The key point is
that the Lagrangian four-form appearing in \eqref{genactform} is the top form of the equivariantly closed form given
in \eqref{eqclosedactwo}.
Thus, we can evaluate using the Berline--Vergne--Atiyah--Bott (BVAB) theorem \cite{BV:1982,Atiyah:1984px}: 
\begin{align}\label{intpsifour}
\int_M\Psi_4^\pm= I_\pm^\text{nuts}  + I_\pm^\text{bolts}+I_{\pm}^{  \partial M}\,.
\end{align} 
Here 
$I_\pm^\text{nuts}$, $I_\pm^\text{bolts}$ are the fixed point contributions, which we assume have no component on the boundary $\partial M$ of the four-manifold $M$, 
and $I_{\pm}^{\partial M}$ is a specific boundary contribution.
To simplify formulae, 
we now assume the fixed point set does not 
have orbifold singularities \footnote{The generalization to orbifolds is straightforward.}.

We henceforth focus on a specific class of theories 
which, after gauge-fixing, are associated with Poincar\'e supergravity theories, both gauged and ungauged, which are
relevant for holography and supersymmetric black holes, as mentioned in the introduction. 
Consider chiral multiplets
that are constructed from 
$n_v+1$ vector multiplets $\mathbb{V}^I$, $I=0,\dots,n_v$, and 
the higher derivative chiral 
Weyl-squared multiplets
$\mathbb{W}^2_\pm$ and 
T-log multiplets 
$\mathbb{T}_\pm$ constructed from $\log\Phiprime_\mp$, where $\Phiprime_\mp$ has 
weight $\wprime$. 
These may be combined into a chiral multiplet of weight $w=2$ 
by introducing 
general real functions $\mathcal{F}^\pm(X^I_\pm, A_\pm|_{\mathbb{W}^2_\pm},A_\pm|_{\mathbb{T}_\pm})$, satisfying 
\begin{align}
&	\lambda^2\mathcal{F}^\pm(X^I_\pm, A_\pm|_{\mathbb{W}^2_\pm},A_\pm|_{\mathbb{T}_\pm})\nonumber\\
 &\ \qquad\qquad  = \mskip2mu \mathcal{F}^\pm(\lambda X^I_\pm, \lambda^2A_\pm|_{\mathbb{W}^2_\pm},\lambda^2 A_\pm|_{\mathbb{T}_\pm})\,.
\end{align}
The bosonic component fields of this chiral multiplet can  be obtained from 
\cite{deWit:2017cle}, 
where $\mathcal{F}^\pm=A_\pm|_{\mathcal{F}^\pm}$ is the lowest weight 
field. 

The  expressions \eqref{eqclosedactwo} show that the 
action for a chiral multiplet only receives contributions from
fixed points where the spinor $\epsilon^i$ has $-$ chirality, 
while the action for an anti-chiral multiplet only receives contributions 
where the spinor $\epsilon^i$ has $+$ chirality. 
We first consider the nut contributions, which can be written in the compact form
\begin{align}\label{genclocresult}
	I_\pm^\text{nuts} 
& =\sum_{\mathrm{nuts}\mp} \pm \frac{16\pi^2}{b_1b_2} 
\nonumber\\
&\ \ \  \times \mathcal{F}^\pm\Big( S_\mp X^I_\pm, 16(b_1\mp b_2)^2, 
\frac{\wprime}{2} (b_1\pm b_2)^2\Big)\, ,
\end{align}
where we 
used \eqref{b1pmb2}. 
For the bolts, which are Riemann surfaces $\Sigma_g$, of genus $g$, we find 
\begin{align}\label{Iboltagain}
		&I^\text{bolts}_\pm = \sum_{\text{bolts$\mp$} } 16 \pi^2 \Bigg[ \mathfrak{p}_{\Sigma_g}^I\partial_{X^I_\pm}\mathcal{F}^\pm \mp \mathcal{F}^\pm \int_{\Sigma_g}
		c_1(N\Sigma_g)\nn\\
		&\ \  \ \ \quad  -32 \frac{\partial\mathcal{F}^\pm}{\partial A_\pm|_{\mathbb{W}^2_\pm}}\Big( \int_{\Sigma_g} c_1(T\Sigma_g) \mp  c_1(N\Sigma_g)\Big)  \nn \\
		&\ \  \ \   \quad 
		+ \wprime \frac{\partial\mathcal{F}^\pm}{\partial A_\pm|_{\mathbb{T}_\pm}}\Big( \int_{\Sigma_g}c_1(T\Sigma_g) \pm c_1(N\Sigma_g)\Big)\Bigg]\, .
\end{align}
Here $\mathcal{F}^\pm$ are evaluated at $\mathcal{F}^\pm(b^{-1}S_\mp X^I_\pm,16,\frac{\wprime}{2})$
where recall $\xi=b\mskip1mu \partial_{\varphi}$ rotates the normal 
direction to the bolt $\Sigma_g$. 
$T\Sigma_g$, $N \Sigma_g$ denote tangent and normal bundles.
Fixing an orientation these are complex line bundles, and $c_1$ denotes the first Chern class. In particular
$\int_{\Sigma_g}c_1(T\Sigma_g)=2-2g$. The magnetic flux $\mathfrak{p}_{\Sigma_g}^I$ 
in \eqref{Iboltagain} is given by
\begin{align}\label{FIWflux}
 \mathfrak{p}^I_{\Sigma_g}=\frac{1}{4\pi}\int_{\Sigma_g} F^I(W)\,,
\end{align}	
where recall $F^I(W)$ is the field strength of the Abelian gauge field in the vector multiplet 
$\mathbb{V}^I$.

The expressions \eqref{genclocresult}, \eqref{Iboltagain} 
depend on chirality signs $\pm$, weights $b_i$, and fluxes/topological invariants, 
but also the scalars $S_\mp X^I_\pm$ evaluated at the fixed points. The former 
is global data, while the latter is not. 
However, the equivariantly  
closed forms $\Psi^I_{(F)}$ given in \eqref{fieldstrengthec} may be used 
to systematically relate the scalars at different fixed points, 
and also to their values on the boundary 
$\partial M$, as explained in 
\cite{BenettiGenolini:2024xeo,BenettiGenolini:2024hyd,BenettiGenolini:2024lbj}.

First consider a two-sphere  $\Sigma=S^2$ \footnote{The extension to $\Sigma$ being a spindle is straightforward.} to which $\xi$ is tangent. If $\xi$ is not identically zero on $\Sigma$ (so $\Sigma$ is not a bolt),
$\xi$ will have fixed points at the north and south poles of the $S^2$, which will be 
two isolated fixed points in the sum in \eqref{genclocresult}. Then from BVAB we deduce
\begin{align}\label{pritstick}
\mathfrak{p}^I_{S^2}=\frac{1}{2b_N}(\Psi^I_{(F)0}|_N-\Psi^I_{(F)0}|_S )\in \mathbb{Z}
\,,
\end{align}
where 
$\Psi^I_{(F)0}= 2(X_+^IS_--X_-^IS_+)$ is the zero-form part of $\Psi^I_{(F)}$ in \eqref{fieldstrengthec}, and we used the fact that the weights at the poles are related by $b_S=-b_N$.
The spinor $\epsilon^i$ will have a fixed $\pm$ chirality at each pole and correspondingly 
$\Psi^I_{(F)0}|_\pm=\mp2 X^I_\mp S_\pm$. Equation \eqref{pritstick}
then determines the scalars at one fixed point in terms of scalars at the other fixed point, together with a magnetic flux $\mathfrak{p}^I_{S^2}$. The topological formula $\chi(M)=1+b_2(M)$, 
valid for simply connected $M$ and in particular for toric $M$, ensures 
there are $\chi(M)-1$ gluing conditions for data at $\chi(M)$ nuts, leaving just one scalar (for each $I$)
undetermined. Here $\chi(M)$, $b_2(M)$ are the Euler and second Betti number of $M$. These are called the
``gluing rules'' (the name is inspired by \cite{Hosseini:2019iad}).

Next consider a non-compact two-dimensional submanifold to which $\xi$ is tangent. 
Assume this has topology $\mathbb{R}^2$ and 
intersects the boundary $\partial M$ on an $S^1$, where we define the holonomy 
$\Delta^I\equiv (4\pi)^{-1}\int_{S^1}W^I=(4\pi)^{-1}\int_{\mathbb{R}^2}F^I(W)$. The origin of the $\mathbb{R}^2$ necessarily lies on the fixed point set of $\xi$, and applying Stokes' and BVAB theorem leads to the relation 
\begin{align}\label{UVIR}
\frac{1}{2b}\Psi^I_{(F)0}|_{\text{fixed point}}=\Delta^I+\ii\frac{2\pi}{b}\sigma^I\, ,
\end{align} 
where $\sigma^I\equiv -{\ii}(4\pi)^{-1}\Psi^I_{(F)0}|_{S^1}$ is the boundary value of $\Psi^I_{(F)0}$. 
These ``UV/IR relations" (see also \cite{Bobev:2020pjk}), combined with the gluing rules, are crucial in the holographic context as they allow one to interpret the action in terms of boundary field theory data 
\cite{BenettiGenolini:2024xeo, BenettiGenolini:2024hyd, BenettiGenolini:2024lbj}. 

Now, to obtain on-shell results for Poincar\'e supergravity, either gauged or ungauged, one needs to add compensating multiplets
and carry out gauge fixing.
Boundary terms need to be carefully taken into account. These include those arising from the BVAB formula \eqref{intpsifour}, 
and in any physical application one will also have
Gibbons--Hawking--York--Myers
boundary terms 
\cite{York:1972sj,Gibbons:1976ue,Myers:1987yn}, as well as boundary terms to remove divergences, either by
implementing holographic renormalization in the gauged case, or background subtraction in the ungauged case, and potentially also a Legendre transform.
We will not analyse these issues further here. However, importantly, 
 in the case of two-derivative on-shell  $D=4$, $\mathcal{N}=2$ gauged supergravity, it has been shown that the combination of all these boundary terms precisely cancel \cite{BenettiGenolini:2024lbj}. 
 Thus, for these two-derivative theories, the holographic on-shell action is 
 exactly given by the fixed point result. 
 Moreover, we have shown that at the two-derivative level, the results in this paper are in precise alignment with those
of \cite{BenettiGenolini:2024lbj} after gauge-fixing.
It is therefore plausible that our fixed point result for off-shell configurations
 given in \eqref{genclocresult}, \eqref{Iboltagain} does indeed give the correct physical result more generally.
 
Our expression for the nut contribution \eqref{genclocresult} was insightfully conjectured 
 in \cite{Hristov:2021qsw,Hristov:2024cgj}, based on extrapolation of known solutions, where gluing rules
 were also discussed. Here we have provided a general proof of \eqref{genclocresult} as well as obtained the bolt contribution \eqref{Iboltagain}. 
 In addition, we have provided a precise prescription for computing the gluing rules in general and, furthermore, 
explained the important UV/IR dictionary. 

\section{ABJM theory}

As an application to holography, we consider the gravity dual of ABJM theory \cite{Aharony:2008ug}. Specifically, we consider $D=4$ STU gauged supergravity with  vector multiplets labelled by $I=0,1,2,3$. Solutions to the two-derivative theory uplift on $S^7/\Z_k$ with $N$ units of seven-form flux
to $D=11$ supergravity \cite{Cvetic:1999xp}, where they are dual to 
the $d=3$, $U(N)\times U(N)$ ABJM theory with Chern--Simons level $k\in \Z$. 
The derivative expansion in gravity corresponds to a $1/N$ expansion in field theory, with 
$k$ fixed. 
It is not known how to compute higher derivative corrections to the $D=4$ prepotential 
directly from M-theory in $D=11$, but an  all-derivative prepotential was 
conjectured in \cite{Hristov:2022lcw}. 
Absorbing the overall Newton constant into the prepotential, in our conventions the latter reads 
\begin{align}\label{STUF}
&\mathcal{F}  = \frac{\sqrt{2k X^0X^1X^2 X^3}}{48\pi}\times \nonumber\\
&   \left[N_k -\frac{1}{3k} \frac{\frac{1}{16}f_{\mathbb{W}^2}(X^I) A|_{\mathbb{W}^2}+\frac{2}{\wwlog'} f_\mathbb{T}(X^I) A|_\mathbb{T}}{16 X^0X^1X^2X^3}\right]^{3/2}\, ,
\end{align}
where $N_k\equiv N - \frac{k}{24}$
and
\begin{align}
f_{\mathbb{W}^2}(X^I) & \equiv -2\sum_{I<J}X^IX^J\, ,\nonumber\\
f_\mathbb{T}(X^I)& \equiv\sum_{I=0}^3 (X^I)^2 - \frac{1}{{\scriptstyle{\sum_{I=0}^3 X^I} }}
(X^0+X^1-X^2-X^3)\times \nonumber\\
&\hskip-0.8cm (X^0-X^1+X^2-X^3)(X^0-X^1-X^2+X^3)\, ,
\end{align}
and in \eqref{FIWflux} we now take $\mathfrak{p}^I_{\Sigma_g}\in\Z$, associated with smooth uplifted solutions.
In the Euclidean theory $\mathcal{F}^\pm$  
may in principle be specified independently, but Wick rotating the
Lorentzian theory 
implies $\mathcal{F}^+=\mathcal{F}^-=\mathcal{F}=\mathcal{F}(X^I,A|_{\mathbb{W}^2},A|_\mathbb{T})$.   
The expression \eqref{STUF} was essentially deduced from field theory 
results on $S^3$, up to a logarithmic term 
and the $N^0$ term in the all-order $1/N$ expansion. The veracity of the conjecture comes from consistency 
with a number of other known results 
\cite{Hristov:2022lcw, Bobev:2022eus}. In particular, 
combining \eqref{STUF} with the ``gluing rules'' for black hole 
solutions with topology $\R^2\times S^2$  gives agreement with 
field theory results for the topologically twisted index (TTI) and superconformal index 
(SCI) of ABJM theory on $S^1\times S^2$. Notice this is a form of consistency 
check of precision holography, rather than a fully independent computation 
of each side of the duality. 

Our bolt formula \eqref{Iboltagain} allows us to present a new result for  
ABJM theory on a general class of supersymmetric three-manifolds. 
The conformal Killing vector $\xi$ is by assumption nowhere zero on the boundary  $\partial M$.
Provided all the orbits of $\xi$ close, $\partial M$ is then a Seifert-fibered three-manifold \footnote{This general class of rigid supersymmetric geometries was studied in  \cite{Closset:2012ru}.}. 
This is the total space of a circle orbibundle over an orbifold Riemann surface $\Sigma_{g,d}$: the Riemann surface $\Sigma_g$ 
has $d$ orbifold points, locally $\mathbb{C}/\mathbb{Z}_{n_i}$, $i=1,\ldots,d$, 
with the circle orbibundle specified by an integer degree $p\in\mathbb{Z}$ 
and parameters $0\leq m_i<n_i$. Such three-manifolds are naturally the boundary 
of the associated complex line orbibundle, with $\xi$ rotating the $\R^2=\mathbb{C}$ fibers. We take this to be our bulk $M$ filling the boundary Seifert manifold, 
which has a bolt  $\Sigma_{g,d}$ at the origin of $\R^2=\mathbb{C}$. Then we may simply substitute 
\eqref{STUF} into~\eqref{Iboltagain}, suitably modified to include orbifold points, with 
\begin{align}
\int_{\Sigma_{g,d}}c_1(T\Sigma_{g,d}) & = 2-2g-d+\sum_{i=1}^d \frac{1}{n_i}\, ,\nonumber\\
\int_{\Sigma_{g,d}}c_1(N\Sigma_{g,d}) & = -p+\sum_{i=1}^d\frac{m_i}{n_i}\, .
\end{align}
$I_\pm^{\text{bolt}}$ then gives an all-order prediction for saddle points of ABJM theory 
on this general class of three-manifolds, where the finite $N$ partition function is
known 
\cite{Closset:2019hyt}. 

While this general large $N$ expansion has not been computed in field theory, 
reference \cite{Hong:2024uns} studied the case that $\Sigma_{g,d}$ is smooth, 
so $d=0$, but with general Chern number $p$ (further setting $p=0$ gives the TTI). The map from gravity variables 
to field theory variables is the same as in the two-derivative theory, given in \cite{BenettiGenolini:2024lbj} 
\begin{align}\label{map}
\text{$-$ chirality} : \quad  b^{-1}S_-X_+^I & = \Delta^I + \ii \frac{2\pi}{b}\sigma^I = 2[\Deltathere^I]\, , \nonumber\\
 \mathfrak{p}^I_{\Sigma_g} & = 2(\mathfrak{n}^I+ pD^I-\tfrac{1}{2}p\nu_R)\, .
\end{align}
Here the first equation is the UV/IR relation \eqref{UVIR}, and the relation 
to the field theory variables $\Deltathere^I$ in \cite{Hong:2024uns}.
Note that the constraint $\sum_{I=0}^3 [\Deltathere^I]=1$
is equivalent to
 $b=\frac{1}{2}\sum_{I=0}^3 S_-X_+^I$ \cite{toappone}. We refer to \cite{Hong:2024uns}
for definitions of all the quantities on the right hand side of \eqref{map}, 
but note  that $\Deltathere^I = [\Deltathere^I]+D^I$ with $D^I$ integers, and 
$\sum_{I=0}^3 \Deltathere^I = 2\nu_R$, where $\nu_R=+\frac{1}{2}$ 
for these negative chirality solutions. 
 It is straightforward to check 
that $I_+^{\text{bolt}}$ given by \eqref{Iboltagain} matches the full $1/N$ 
expansion given in equation (3.25) of \cite{Hong:2024uns} (up to 
the logarithmic term and $N^0$ term), and that the positive chirality branch $I_-^{\text{bolt}}$
matches the second saddle point discussed in \cite{Hong:2024uns}. 
This is a remarkable agreement and provides non-trivial  support for
\eqref{STUF}, 
as well as our general expectation that 
in evaluating the higher derivative actions using localization, the total boundary contributions will vanish. 

\section*{Acknowledgments}
We thank Alice L\"uscher for discussions. 
This work was supported in part by STFC grants ST/X000575/1 and
ST/X000761/1, EPSRC grant EP/R014604/1, and SNSF Ambizione grant PZ00P2\_208666.
FG is supported by an STFC studentship.


%

\appendix

\section{Supplementary material: }

We record here that the lowest weight fields of the T-log multiplets, ${\mathbb T}_\pm$, expressed in terms of
the components of $\Phiprime_\mp$ are given by
\begin{equation}\label{tlogaexpressions}
		A_\pm|_{\mathbb{T}_\pm}
		=\frac{C_\mp}{A_\mp}+\frac{1}{4A_\mp^2}\Big[ 2F^{\pm ab}F^\pm_{ab}-\varepsilon_{ij}\varepsilon_{kl}B^{ik}_{\mp}B^{jl}_{\mp}\Big]\,.
\end{equation}

A key equation in our analysis is an expression for the two-form $\dd\xi^\flat$, which we
can decompose 
into self-dual and anti-self dual components, $\dd\xi^\flat=(\dd\xi^\flat)^++
(\dd\xi^\flat)^-$.
Using (anti)-chiral multiplets $\Phiprime_\pm$ of \emph{arbitrary} weight $\wprime$ we 
find 
\begin{align}\label{dxieqappendix}
		(\dd\xi^\flat)^\pm&=\mp\frac{1}{2}S_\pm T^\pm\pm\frac{1}{\wprime A_\mp}\Big[2S_\mp F^\pm+\nn\\
&\qquad\qquad		B^{ij}_\mp\epsilon_{jk}\newOmega_{\mp i}{}^k \mp2 ( DA_\mp \wedge \xi^\flat )^\pm\Big]\,.
\end{align}
This follows from imposing supersymmetry for the Weyl multiplet and 
a chiral multiplet of weight $\wprime$, with components \eqref{chiralcomponents}. The covariant derivative acting on $A$ is 
$D_\mu A_\pm = (\partial_\mu 
\pm \wprime A_\mu)A_\pm$. 
Evaluating the norms of these two-forms at a fixed point of chirality $\pm$ leads to~\eqref{b1pmb2}.

\end{document}